\documentclass[lettersize,journal]{IEEEtran}
\usepackage{amsmath,amsfonts}
\usepackage{algorithmic}
\usepackage{algorithm}
\usepackage{array}
\usepackage[caption=false,font=normalsize,labelfont=sf,textfont=sf]{subfig}
\usepackage{textcomp}
\usepackage{stfloats}
\usepackage{url}
\usepackage{verbatim}
\usepackage{graphicx}
\usepackage{cite}

\usepackage{amsmath,amssymb}
\usepackage{amsfonts}

\usepackage{tabularx}
\usepackage{caption}

\usepackage{booktabs}
\usepackage{ragged2e}

\usepackage{booktabs}
\usepackage{tabularx}
\usepackage{multirow}
\usepackage{ragged2e}

\usepackage{subcaption}

\newcolumntype{Y}{>{\RaggedRight\arraybackslash}X}

\begin{document}



 \title{A Parameter-Specific Retrieval and Knowledge-Guided Reasoning Framework for LLM-Based GPSR Optimization in FANETs}


\author{Zhipeng Lin, Bin Duo, \IEEEmembership{Member,~IEEE,} Tong Liu, Jie Lin, Jianting Yuan and Xiaojun Yuan, \IEEEmembership{Fellow,~IEEE}

\thanks{\textit{(Corresponding author: Bin Duo.)}}
\thanks{Zhipeng Lin, Bin Duo, Tong Liu, Jie Lin are with the College of Computer Science and Cyber Security, Chengdu University of Technology, Chengdu 610059, China (e-mail: zhipenglin02@163.com; duobin@cdut.edu.cn; liutong1982@cdut.edu.cn; linjie07@cdut.edu.cn).}
\thanks{Jianting Yuan is with the School of Software, Xinjiang University, Urumqi 830046, China (e-mail: steven@xju.edu.cn).}
\thanks{Xiaojun Yuan is with the National Key Laboratory of Wireless Communications, University of Electronic Science and Technology of China, Chengdu 611731, China (e-mail: xjyuan@uestc.edu.cn).}

}

\markboth{}%
{Shell \MakeLowercase{\textit{et al.}}: A Sample Article Using IEEEtran.cls for IEEE Journals}

\maketitle

\begin{abstract}
Existing Greedy Perimeter Stateless Routing (GPSR)-based protocols for Flying Ad-Hoc Networks (FANETs) struggle to adapt routing parameters, such as hello interval, multi-path number, and greedy forwarding weights, under highly dynamic environments. As an emerging artificial intelligence technology, large language models (LLMs) show potential for intelligent decision-making, providing new opportunities for adaptive adjustment of GPSR parameters to improve network performance. However, applying LLMs to GPSR remains challenging due to irrelevant experience retrieval and the absence of protocol constraints. To address these issues, we propose a Parameter-Specific Multi-Index Retrieval and Knowledge-Guided Reasoning framework for adaptive GPSR optimization (PMKR-GPSR), an LLM-based framework that enables protocol-consistent routing parameter adaptation. We design a parameter-specific multi-index retrieval mechanism to provide LLMs with parameter-relevant experiences while reducing interference from irrelevant information. We further construct a knowledge-guided constraint graph to enforce that the routing parameters satisfy dependency rules and optimization constraints. Simulation results demonstrate that PMKR-GPSR achieves higher packet delivery ratio and lower end-to-end delay under high-mobility FANETs.

\end{abstract}


\section{Introduction}
\IEEEPARstart{F}{lying} Ad Hoc Networks (FANETs) have attracted considerable attention due to their rapid deployment capability and flexible topology. Compared to conventional mobile ad hoc networks, FANETs are characterized by rapidly changing topologies and unstable wireless links, significantly increasing the complexity of routing decisions \cite{multi-path}. Consequently, designing adaptive routing mechanisms capable of maintaining reliable communication remains a critical challenge.

Among existing geographic routing protocols, Greedy Perimeter Stateless Routing (GPSR) has been widely adopted due to its low computational complexity and localized forwarding strategy. However, conventional GPSR employs fixed forwarding strategies. In highly dynamic FANETs, fixed routing parameters struggle to adapt to diverse communication scenarios, resulting in suboptimal forwarding decisions, unstable routes, and degraded packet delivery performance\cite{GPSR_Prob}.

To improve GPSR adaptability, numerous optimization methods have been proposed. Some studies enhanced greedy forwarding by considering multiple routing metrics, enabling a more comprehensive evaluation of forwarding nodes and improving forwarding reliability \cite{CFGPSR}. Other approaches adopted reinforcement learning (RL) to learn adaptive routing strategies through continuous interaction with the network environment \cite{QLGR}. However, RL-based methods generally require extensive exploration and carefully designed reward functions to achieve stable convergence\cite{RL_Prob}. As network topologies evolve, previously learned policies may become suboptimal, requiring additional adaptation to maintain routing performance in highly dynamic scenarios.

Recently, large language models (LLMs) have demonstrated potential for intelligent decision-making in wireless communications due to their knowledge utilization and reasoning capabilities  \cite{LLM_Prob}. However, existing studies have shown that LLMs without explicit knowledge constraints and relevant experience may generate hallucinated or infeasible outputs \cite{kc_2}. 


Therefore, directly applying LLMs to GPSR optimization remains challenging. First, different routing parameters, such as hello interval, multi-path number, and greedy forwarding weights, have heterogeneous dependencies on different network features, causing irrelevant information to interfere with LLM reasoning. Second, the lack of protocol knowledge constraints prevents LLMs from understanding the relationships among network conditions, routing parameters, and performance objectives, which may lead to unreasonable parameter adjustments that violate GPSR principles and deteriorate routing performance. 

To address these challenges, we propose a Parameter-Specific Multi-Index Retrieval and Knowledge-Guided Reasoning framework for adaptive GPSR optimization (PMKR-GPSR). First, to address the heterogeneous feature dependencies of different routing parameters and the interference caused by unified experience retrieval, we design a parameter-specific multi-index retrieval mechanism. Network features are extracted to query corresponding parameter-specific indexes, providing the LLM with relevant experiences for routing parameter generation. Second, to address the lack of protocol knowledge constraints, we construct a knowledge-guided constraint graph. The graph incorporates parameter dependency rules and optimization constraints for LLM reasoning, guiding the generation of routing parameters. Simulation results demonstrate that PMKR-GPSR achieves improved packet delivery ratio (PDR) and reduced end-to-end (E2E) delay under high-mobility FANET scenarios, validating its effectiveness in enhancing routing adaptability and robustness in dynamic network environments. 

\begin{figure*}[t!] 
   \centering
   \includegraphics[width=0.6\linewidth]{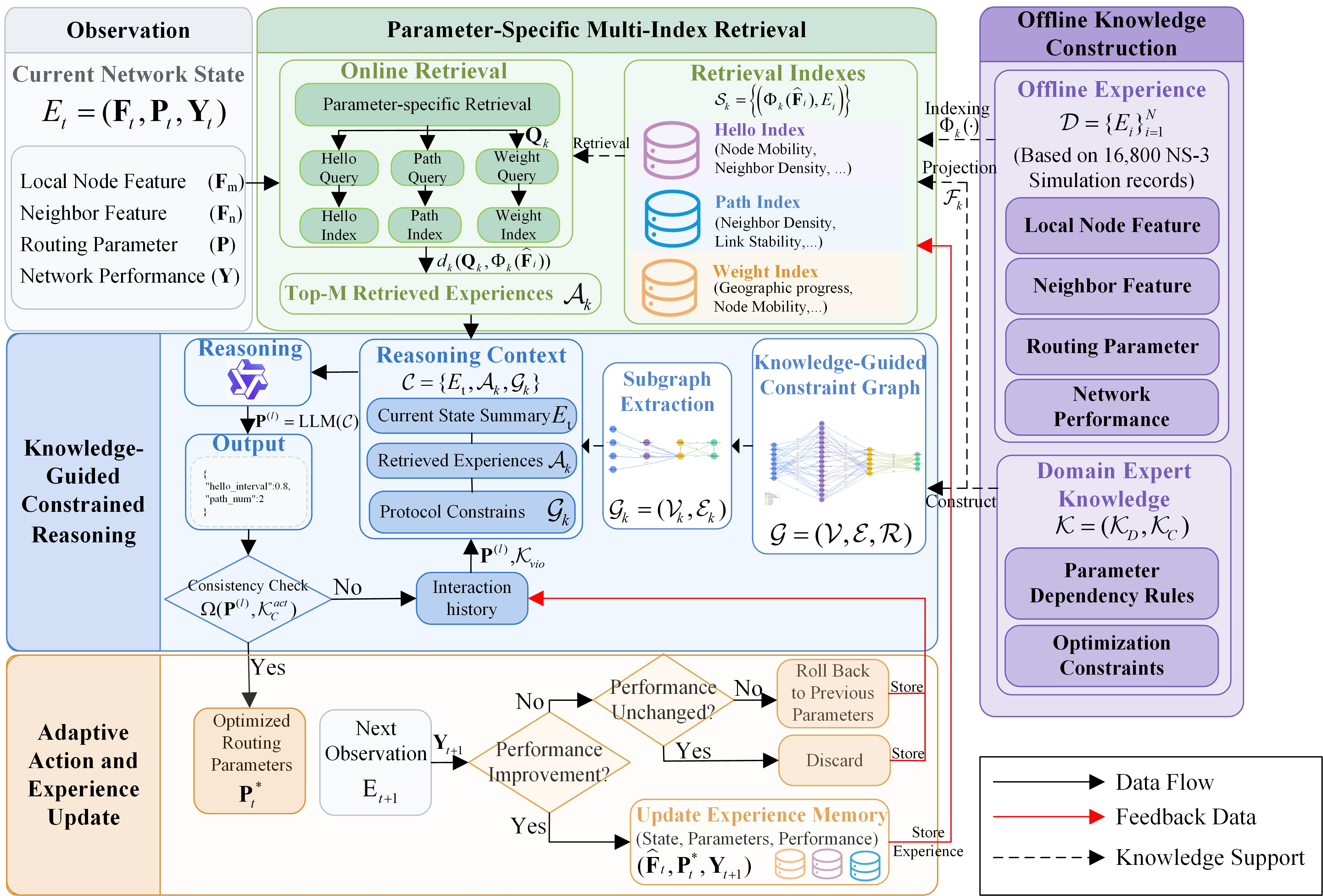}
   \caption{Architecture of the Proposed PMKR-GPSR Framework }
   \label{fig:Architecture_picture}
   \vspace{-6pt}
\end{figure*}

\section{Problem Formulation}
\subsection{UAV Network Model  }
We consider a highly dynamic UAV ad hoc network, modeled as an undirected  graph \(G=(U,L)\), where \(U=\{ u_i \mid 1 \le i \le N \}\) denotes the set of \(N\) UAVs and \(L\subseteq U\times U\) represents the communication links established between neighboring UAVs within the transmission range. 

\subsection{GPSR Routing Protocol}
The proposed framework is built on the GPSR protocol \cite{GPSR}. Each node periodically broadcasts Hello packets and updates its neighbor table by receiving Hello packets from one-hop neighbors. Based on the neighbor table, the forwarding node selects the neighbor closest to the destination for packet forwarding. 

Following existing studies on GPSR optimization, this work adopts an adaptive Hello strategy and a multi-path forwarding strategy to improve transmission reliability during packet delivery \cite{multi-path}. In addition, a multi-parameter greedy forwarding strategy is employed to enhance the accuracy of next-hop selection \cite{GPSR_Prob}. The forwarding priority of neighbor node \(u_j\) is calculated as 
\begin{equation}
    S_j=w_d\hat{D}_j+w_l\hat{L}_j+w_v\hat{V}_j+w_n\hat{N}_j,
    \label{forwarding_strategy}
\end{equation}
where \(\hat{D}_j\), \(\hat{L}_j\), \(\hat{V}_j\), and \(\hat{N}_j\) denote the normalized distance to destination, link lifetime, relative velocity, and neighbor degree of candidate node \(u_j\), with their corresponding weights \(w_d\), \(w_l\), \(w_v\) and \(w_n\) satisfying \(w_i \ge 0, \sum_i w_i = 1\).

\begin{table*}[t]
\centering
\caption{Parameter Dependency Rules for Adaptive GPSR Optimization.}
\label{tab1}
\vspace{-6pt}
\begin{tabularx}{\textwidth}{l l l Y}
\toprule
\textbf{Routing Parameter} & \textbf{Scenario Feature} & \textbf{Key Metrics} & \textbf{Rationale} \\
\midrule

\multirow{2}{*}{Hello Interval}
& Relative mobility
& $v_{\text{self}}, \mu(v_{\text{rel}}), \sigma(v_{\text{rel}})$
& Topology changes induced by high mobility necessitate shorter hello intervals. \\
\cmidrule(lr){2-4}
& Energy condition
& $e$
& Low-energy nodes should reduce Hello overhead to preserve network lifetime. \\
\midrule

\multirow{3}{*}{Path Number}
& Forwarding candidates
& $c_c$
& Abundant forwarding candidates enable additional disjoint paths. \\
\cmidrule(lr){2-4}
& Link stability
& $t_h^{*}, \mu(t_h), \sigma(t_h)$
& Frequent link failures require more paths to improve reliability. \\
\cmidrule(lr){2-4}
& Network congestion
& $q_{\text{self}}, q_{\min}, \mu(q)$
& Congested nodes should restrict path counts to reduce overhead. \\
\midrule

\multirow{3}{*}{Forwarding Weight}
& Geographic progress
& $r_p^{*}, \mu(r_p)$
& Maximizes geographic progress toward the destination. \\
\cmidrule(lr){2-4}
& Relative mobility
& $\mu(v_{\text{rel}}), \sigma(v_{\text{rel}})$
& Focuses on relative velocity and link lifetime under high mobility. \\
\cmidrule(lr){2-4}
& Neighbor connectivity
& $n_c, \mu(n_c)$
& Neighbor sets mitigate routing void risks emerging in sparse regions. \\
\bottomrule
\end{tabularx}

\vspace{-10pt}
\end{table*}

\section{Proposed PMKR-GPSR Framework }

Fig. \ref{fig:Architecture_picture} illustrates the overall PMKR-GPSR framework. In the offline stage, routing experiences are organized into parameter-specific retrieval indexes, and domain knowledge is encoded into a knowledge-guided constraint graph. During online optimization, the framework retrieves relevant experiences, performs knowledge-guided reasoning, generates adaptive routing parameters, and updates experiences through feedback. The proposed framework enhances the reasoning process of an LLM rather than fine-tuning its model parameters. This design avoids the need for large-scale routing-specific training while enabling adaptation to unseen network scenarios. Moreover, PMKR-GPSR explicitly incorporates protocol knowledge into the reasoning process, thereby improving the reliability and interpretability of the generated routing decisions.

\subsection{Offline Knowledge Construction}
\subsubsection{Domain Expert Knowledge}
Domain expert knowledge is represented as \(\mathcal{K}=(\mathcal{K}_{D},\mathcal{K}_{C})\), where $\mathcal{K}_{D}$ denotes the parameter dependency rules and \(\mathcal{K}_{C}\) denotes the optimization constraints. Parameter dependency rules specify the scenario features relevant to each routing parameter. Formally, \(\mathcal{K}_{D}=\left\{(p_k,\mathcal{F}_k)\mid p_k\in\mathbf{P}\right\}\), where \(\mathbf{P}=\{H,M,\mathbf{w}\}\) denotes the set of routing parameters, \(H\) represents the hello interval, \(M\) denotes the number of paths, and \(\mathbf{w}=[w_d,w_l,w_v,w_n]\) represents the weights vector of the multi-parameter greedy forwarding strategy in Eq.(\ref{forwarding_strategy}). \(\mathcal{F}_k\) denotes the scenario feature subset relevant to the optimization of routing parameter \(p_k\), which is determined according to the parameter  dependency rules summarized in Table \ref{tab1}. 

Table \ref{tab1} summarizes the parameter dependency rules for routing parameter optimization. The hello interval is determined by relative mobility and energy condition, balancing neighborhood freshness and communication overhead \cite{hello_index}. The number of forwarding paths is associated with forwarding candidates, link stability, and network congestion to improve routing reliability while limiting unnecessary transmissions \cite{path_number_index}. The forwarding weight mainly depends on geographic progress, relative mobility, and neighbor connectivity, enabling the protocol to balance forwarding efficiency, link stability, and robustness under different network conditions \cite{weight_index}. 

Optimization constraints specify the feasible parameter adaptation rules. They are represented as \(\mathcal{K}_{C}=\{\psi_j\}_{j=1}^{N_c}\), where \(N_c\) denotes the number of constraint rules. Each constraint is formulated as \(\psi_j:\Gamma_j\rightarrow\Delta_j\), where \(\Gamma_j\) denotes the network state condition defined based on the scenario features, and \(\Delta_j\) represents the executable parameter adaptation constraints, including allowable ranges and adjustment directions. The constraint knowledge is used for knowledge-guided reasoning during online optimization.

\subsubsection{Offline Experience Representation}
This paper constructs an offline experience database \(\mathcal{D}\) from simulation records: \(\mathcal{D}=\{E_i\}_{i=1}^{N}\), where \(N\) denotes the total number of experiences and \(i\) is the experience index. Each experience is represented as \(E=(\mathbf{F},\mathbf{P},\mathbf{Y})\), where \(\mathbf{F}=[\mathbf{F}_{s},\mathbf{F}_{n}]^\top\) denotes the scenario feature representation. \(\mathbf{F}_{s}\) represents self-node information, including node speed \(v_\text{self}\), candidate count \(c_{c}\), queue length \(q_\text{self}\) and energy \(e\). Here, the candidate nodes refer to neighboring nodes that are geographically closer to the destination than the current node. \(\mathbf{F}_{n}\) describes neighboring-node information, including the number of neighbors \(n_c\), geographic progress ratio \(r_p\), relative speed \(v_\text{rel}\), link lifetime \(t_h\), and neighbor queue length \(q\). The geographic progress ratio \(r_p\) represents the relative percentage of geographic advancement toward the destination achieved by a candidate forwarding node. To address the variable number of neighboring nodes and obtain a fixed-dimensional representation of neighbor node information, each neighbor attribute is summarized using statistical descriptors. According to attribute characteristics, different combinations of the best candidate (*), minimum (min), mean (\(\mu\)), and standard deviation (\(\sigma\)) are adopted. The performance vector is defined as \(\mathbf{Y}=\{y_1,y_2,y_3,y_4\}\), including packet delivery ratio, average end-to-end delay, control overhead, and energy consumption. 

Since \(\mathbf{F}\) have heterogeneous scales, we apply Z-score normalization before index construction. The normalization statistics \(\{\mu_l^{norm},\sigma_l^{norm}\}\) are computed from the offline dataset and remain fixed during online inference. Each feature dimension is normalized as 
\begin{equation}
z_l = \frac{x_l - \mu_l^{norm}}{\sigma_l^{norm}},
    \label{Normalization}
\end{equation}
where \(x_l\) represents the raw value of the \textit{l}-th feature,  \(\mu_l^{norm}\) and \(\sigma_l^{norm}\) denote the offline mean and standard deviation of the \textit{l}-th feature. The normalized feature is denoted as \(\hat{\mathbf F}\). 

\subsection{Parameter-Specific Multi-Index Retrieval}
Due to heterogeneous dependencies between routing parameters and network states, unified retrieval may introduce irrelevant interference and degrade experience matching. Therefore, parameter-specific retrieval indexes are constructed based on parameter dependency rules \(\mathcal{K}_D\) to improve retrieval relevance.

Specifically, the feature mapping function \(\Phi_k(\cdot)\) is applied to extract the parameter-specific feature representation from the scenario feature vector: \(\Phi_k(\hat{\mathbf{F}}_i)=\hat{\mathbf{F}}_{i,\mathcal{F}_k}\), where \(\hat{\mathbf{F}}_{i,\mathcal{F}_k}\) denotes the feature subset selected according to \(\mathcal{K}_{D}\). The extracted representation is stored together with the corresponding experience \(E_i\) to construct the parameter-specific retrieval index:
\begin{equation}
\mathcal{S}_k=
\left\{
\left(\Phi_k(\hat{\mathbf{F}}_i),E_i\right)
\mid
E_i=(\mathbf{F}_i,\mathbf{P}_i,\mathbf{Y}_i)\in\mathcal D
\right\},
\label{experience-projected}
\end{equation}
where \(\mathcal{S}_k\) denotes the retrieval index for parameter \(p_k\).

During online optimization, the observation module monitors the current network state \(E_t\) at time \(t\) and constructs the scenario representation \(\mathbf{F}_t\). After normalization in Eq.(\ref{Normalization}), \(\hat{\mathbf{F}}_t\) is projected by \(\Phi_k(\cdot)\) to obtain the query representation \(\mathbf{Q}_k=\Phi_k(\hat{\mathbf{F}}_t)\). The similarity between the query representation \(\mathbf{Q}_k\) and each projected feature representation in the corresponding parameter-specific retrieval index \(\mathcal S_k\) is computed as
\begin{equation}
d_k(\mathbf{Q}_k,\Phi_k(\hat{\mathbf{F}}_i))
=
-\left\|\mathbf{Q}_k-\Phi_k(\hat{\mathbf{F}}_i)\right\|_2,
\label{similar-retrieved}
\end{equation}
where \(d_k(\cdot,\cdot)\) denotes the similarity function. The retrieved experience set is defined as the Top-\(M\) experiences with the highest similarity scores:
\begin{equation}
\mathcal{A}_k = \left\{ E_i \,\bigg|\, \big(\Phi_k(\hat{\mathbf{F}}_i), E_i\big) \in \operatorname{Top}_M\big(\mathcal{S}_k, d_k\big(\mathbf{Q}_k, \cdot\big)\big) \right\},
\label{retrieved-experience-set}
\end{equation}
where \(\mathcal A_k\subseteq\mathcal D\) contains the Top-\(M\) most similar historical experiences for optimizing parameter \(p_k\). Here, \(\operatorname{Top}_M(\mathcal S_k, d_k(\mathbf{Q}_k,\cdot))\) selects the \(M\) retrieval entries in \(\mathcal S_k\) with the highest similarity scores computed by  (\ref{similar-retrieved}).

The retrieved experience set provides parameter-specific knowledge for reasoning and routing parameter optimization. 

\subsection{Knowledge-Guided Constrained Reasoning   }
Although multi-index retrieval provides parameter-relevant experiences, it lacks explicit protocol constraints. Therefore, the domain expert knowledge \(\mathcal K\) is encoded into a knowledge-guided constraint graph to guide LLM reasoning.

The constraint graph is modeled as a directed heterogeneous multi-relational graph, formally represented as \(\mathcal{G}=(\mathcal{V},\mathcal{E},\mathcal{R})\). The node set \(\mathcal{V}=\mathcal{V}_{s}\cup\mathcal{V}_{pr}\cup\mathcal{V}_{p}\cup\mathcal{V}_{m}\) contains network state, routing principle, routing parameter, and performance metric entities. The dependency rules in  \(\mathcal{K}_{D}\) and the constraints in \(\mathcal{K}_{C}\) are mapped into two semantic relation types: Influence (\(r_{I}\)) and Constraint (\(r_{C}\)), defining the relation set \(\mathcal{R}=\{r_{I},r_{C}\}\). Accordingly, the edge set is represented as \(\mathcal{E}=\mathcal{E}_{I}\cup\mathcal{E}_{C}\), where \(\mathcal{E}_{I}\) and \(\mathcal{E}_{C}\) denote the sets of edges instantiated by the Influence and Constraint relations, respectively. Each semantic edge is formulated as a relational triplet \(e=(v_{i},r,v_{j})\), where \(e\in\mathcal{E}\), \(v_{i},v_{j}\in\mathcal{V}\), and \(r\in\mathcal{R}\).

In the constraint graph, each relation type is represented by an adjacency matrix: \(\mathbf{A}^{(r)}\in\{0,1\}^{|\mathcal{V}|\times|\mathcal{V}|}\), where \(\mathbf{A}^{(r)}_{ij}=1\) indicates the existence of a relation \(r\) from \(v_i\) to \(v_j\).

During online optimization, a parameter-specific subgraph is extracted by traversing the constraint graph from the target parameter entity \(v_{p_k}\). For each relation type \(r\), the neighboring entities are obtained as
\begin{equation}
\mathcal{N}_k^{(r)}=\{v_j \mid \mathbf{A}^{(r)}_{ji}=1,\, v_i=v_{p_k}\}.
\label{neighboring-entities}
\end{equation}
The node set of the parameter-specific subgraph is:
\begin{equation}
\mathcal{V}_k=\{v_{p_k}\}\cup\bigcup_{r\in\mathcal{R}}\mathcal{N}_k^{(r)},
\label{parameter-specific subgraph}
\end{equation}
and the corresponding edge set is defined as \(\mathcal{E}_k=\{(v_i,r,v_j)\in\mathcal{E}\mid v_i,v_j\in\mathcal{V}_k\}\). Therefore, the retrieved subgraph is represented as \(\mathcal{G}_k=(\mathcal{V}_k,\mathcal{E}_k)\).

The LLM performs routing optimization based on an integrated reasoning context: \(\mathcal{C}=\{E_{t},\mathcal{A}_k,\mathcal{G}_k\}\). The integrated reasoning context is provided to the LLM, which performs constrained reasoning to generate a routing parameters: \(\mathbf{P}^{(0)}=\operatorname{LLM}(\mathcal{C})\) where \(\mathbf{P}^{(0)}\) represents the initial routing parameters generated. During the iterative refinement process, $\mathbf{P}^{(l)}$ represents the routing parameter set at the $l$-th iteration.

The activated constraint set is obtained as:
\begin{equation}
\mathcal K_C^{act}
=
\{\psi_j\in\mathcal K_C
\mid
\mathbf{F}_{t}\models \Gamma_j
\}.
\label{executable-rules}
\end{equation}

The consistency score is defined as:
\begin{equation}
\Omega(\mathbf{P}^{(l)},\mathcal K_C^{act})
=\frac{1}{|\mathcal K_C^{act}|}\sum_{\psi_j\in\mathcal K_C^{act}}\mathbb{I}(\mathbf{P}^{(l)}\models\psi_j),
\label{consistency-score}
\end{equation}
where \(\mathbb{I}(\cdot)\) indicates whether the generated configuration satisfies the corresponding constraint. It is defined as:
\begin{equation}
 \mathbb{I}(\mathbf{P}^{(l)} \models \psi_j) =
\begin{cases}
1, & \text{if } \mathbf{P}^{(l)} \models \psi_j,  \\
0, & \text{otherwise}.
\end{cases}
\label{generated configuration satisfies}
\end{equation}
If \(\Omega(\mathbf{P}^{(l)},\mathcal K_C^{act})<1\), the violated constraints are summarized into a violation set \(\mathcal{K}_{vio}=\{\psi_j\in\mathcal K_C^{act}\mid \mathbf{P}^{(l)}\not\models \psi_j\}\). \(\mathcal{K}_{vio}\) are incorporated into the interaction history and provided to the LLM as additional reasoning context for iterative refinement:
\begin{equation}
\mathbf{P}^{(l+1)}=\operatorname{LLM}(\mathcal{C},\mathbf{P}^{(l)},\mathcal{K}_{vio}),
\label{LLM-iteratively}
\end{equation}
until all active constraints are satisfied or the maximum number of iterations \(L_{max}\) is reached. The final routing parameters are obtained as: \(\mathbf{P}_t^{\star}=\mathbf{P}^{(l)}.\)

Consequently, the reasoning process is constrained to protocol-consistent routing configurations, improving the reliability of adaptive routing decisions.

\subsection{Adaptive Action and Experience Update}
After knowledge-guided constrained reasoning, the optimized routing parameters \(\mathbf{P}_t^*\) are adopted as adaptive actions. The subsequent network response is monitored to form a new experience: \(E_{t+1}\).

To maintain high-quality experiences in the experience database \(\mathcal{D}\), the newly optimized routing parameters are evaluated based on the subsequent routing performance \(Y_{t+1}\). If the performance is improved, the experience generated from the previous normalized scenario representation \(\hat{\mathbf{F}}_{t}\), optimized configuration \(\mathbf{P}_{t}^{*}\), and performance \(\mathbf{Y}_{t+1}\) is retained:
\begin{equation}
\begin{aligned}
\mathcal{D} &\leftarrow \mathcal{D}\cup\{(\hat{\mathbf{F}}_{t},\mathbf{P}_{t}^{*},\mathbf{Y}_{t+1})\}.
\end{aligned}
\end{equation}
Specifically, parameter-specific features are extracted using the projection operators \(\Phi_{k}(\cdot)\), and the corresponding retrieval indexes are updated with the new experience:
\begin{equation}
\mathcal{S}_{k}\leftarrow\mathcal{S}_{k}\cup
\{(\Phi_{k}(\hat{\mathbf{F}}_{t}),(\hat{\mathbf{F}}_{t},\mathbf{P}_{t}^{*},\mathbf{Y}_{t+1}))\}.
\end{equation}
Otherwise, if no performance improvement is observed, the new configuration is discarded or the system rolls back to the previous routing parameters. The latest state, action, and performance are stored in the interaction history to provide temporal context for future optimization decisions rather than an additional reasoning loop.

\section{Performance Evaluation }
The proposed framework is evaluated using the NS-3.34 network simulator. A three-dimensional FANET with 200 UAVs is deployed in a \(2000\times2000\times200\ m^3\) area, where the communication range is set to 250 m . The UAV mobility follows the Gaussian-Markov mobility model, and the node speed varies from 10 to 80 m/s. Each simulation lasts for 600 s. The traffic model is the constant bit rate model (CBR) with a bit rate equal to 2 Mbps, and each data packet is 1024 bytes. The proposed optimization framework is triggered every 10 s on the source node and forwarding nodes. The offline experience database contains \(N=16,800\) routing experiences, with \(M=5\) retrieved experiences and \(L_{max}=3\) maximum LLM refinement iterations.


The proposed framework is compared with GPSR (original geographic protocol) \cite{GPSR}, CF-GPSR (multi-metric geographic  protocol) \cite{CFGPSR}, and QLGR (reinforcement learning-based protocol) \cite{QLGR}. To evaluate component contributions, the full model and four ablation variants are evaluated: w/o Multi-Index, w/o KG, w/o Multi-Index \& KG, and LLM-only.  Specifically, w/o Multi-Index replaces parameter-specific retrieval indexes with a unified index; w/o KG removes the knowledge-guided constraint graph; w/o Multi-Index \& KG removes knowledge-guided constraints from the single-index retrieval; and LLM-only directly uses the LLM for routing parameters generation without retrieval or knowledge constraints. 

Table \ref{tab:Inference_latencyr} evaluates the inference latency and routing performance of different Qwen2.5-Instruct model scales under a UAV speed of 40 m/s. The results show that smaller models (0.5B and 1.5B) exhibit limited reasoning capability for analyzing routing parameter dependencies and protocol constraints, resulting in insufficient optimization effectiveness and performance close to the original GPSR. Although 3B and 7B models achieve better routing performance, the 7B model introduces substantially higher inference latency with limited performance improvement. Considering the trade-off between reasoning capability and computational overhead, Qwen2.5-3B-Instruct is selected as the simulation model.

\begin{table}
    \centering
\caption{Performance evaluation of different model scales. }
\label{tab:Inference_latencyr}
    \begin{tabular}{|c|c|c|c|c|}\hline
         Model Scales&  0.5B&  1.5B&  3B& 7B\\\hline
         PDR (\%)&  35.6101&  37.4435&  67.8925& 68.4435
\\ \hline
 Delay (ms)& 326.8720& 277.3431& 232.1994&238.2675\\\hline
 Inference latency (ms)& 1838.770& 2126.063& 2661.467&3664.793\\\hline
    \end{tabular}
    
    \vspace{-10pt}
\end{table}

\begin{figure*}[t] 
    \centering
    \begin{minipage}[t]{0.24\linewidth}
        \centering
        \includegraphics[width=\linewidth]{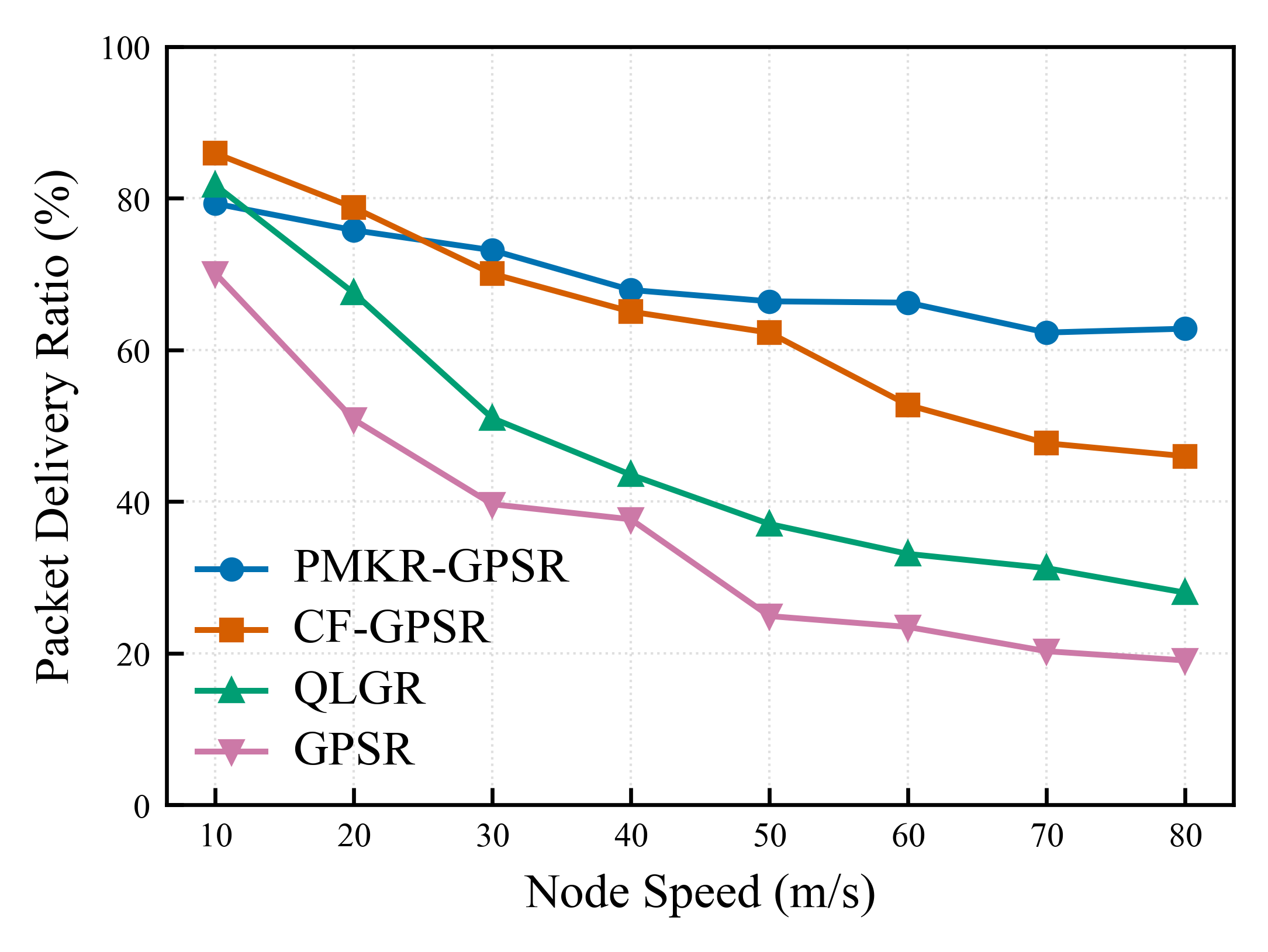}
        \caption{PDR: Baselines.}
        \label{fig:pdr_vs_speed_baselines}
    \end{minipage}
    \hfill
    \begin{minipage}[t]{0.24\linewidth}
        \centering
        \includegraphics[width=\linewidth]{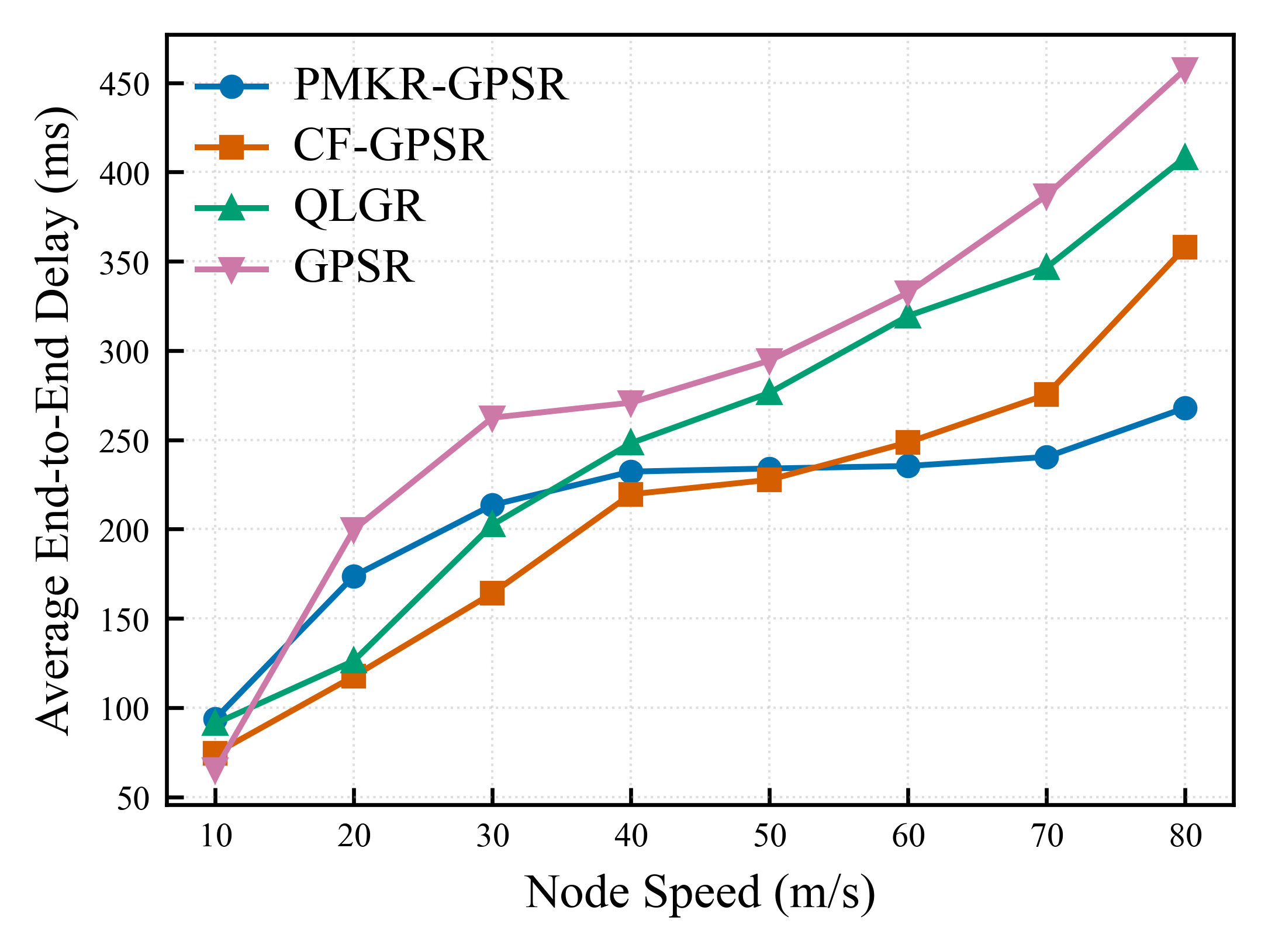}
        \caption{Delay: Baselines}
        \label{fig:delay_vs_speed_baselines}
    \end{minipage}
    \hfill
    \begin{minipage}[t]{0.24\linewidth}
        \centering
        \includegraphics[width=\linewidth]{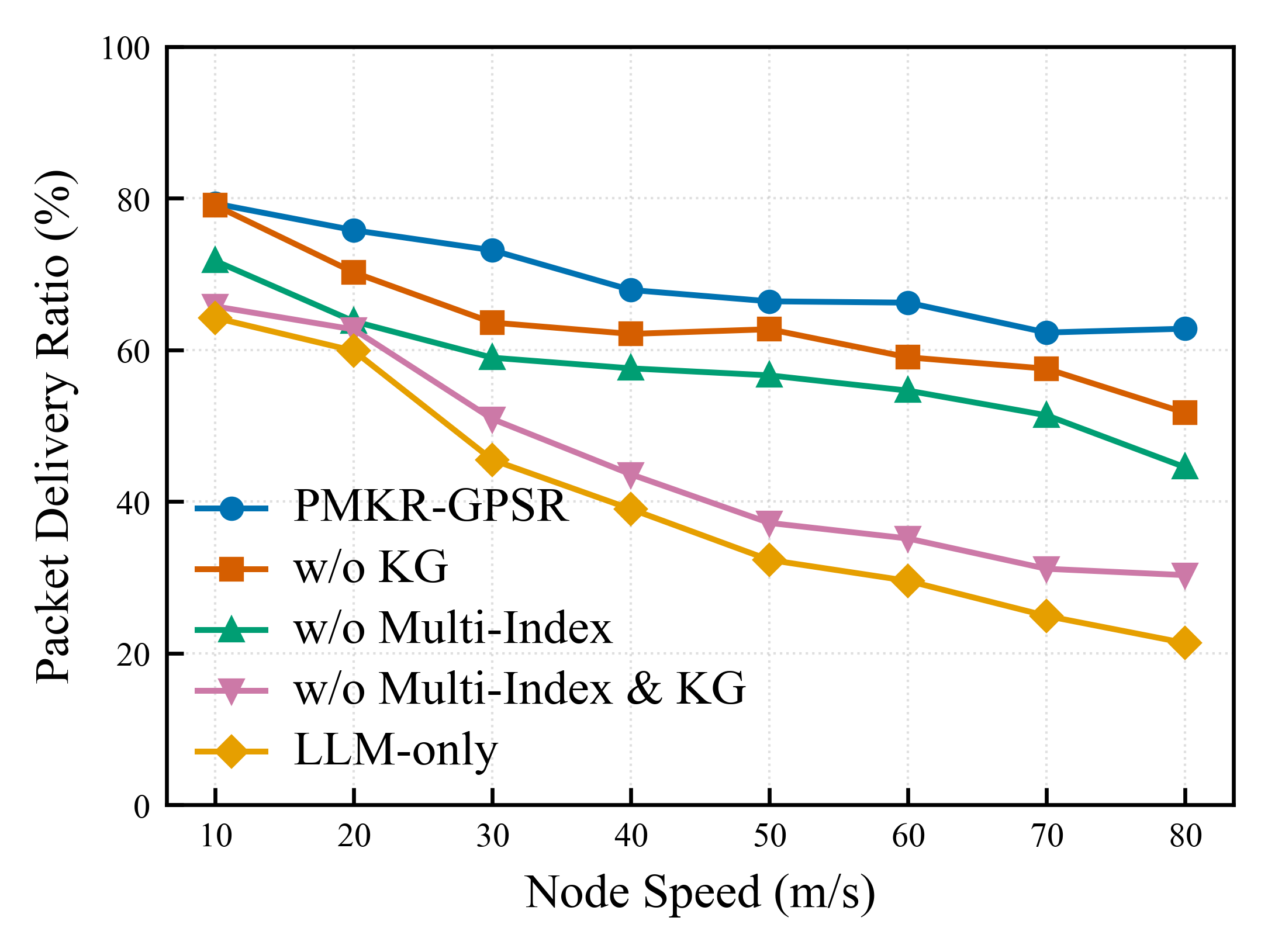}
        \caption{PDR: Ablation.}
        \label{fig:pdr_vs_speed_index_ablation}
    \end{minipage}
    \hfill
    \begin{minipage}[t]{0.24\linewidth}
        \centering
        \includegraphics[width=\linewidth]{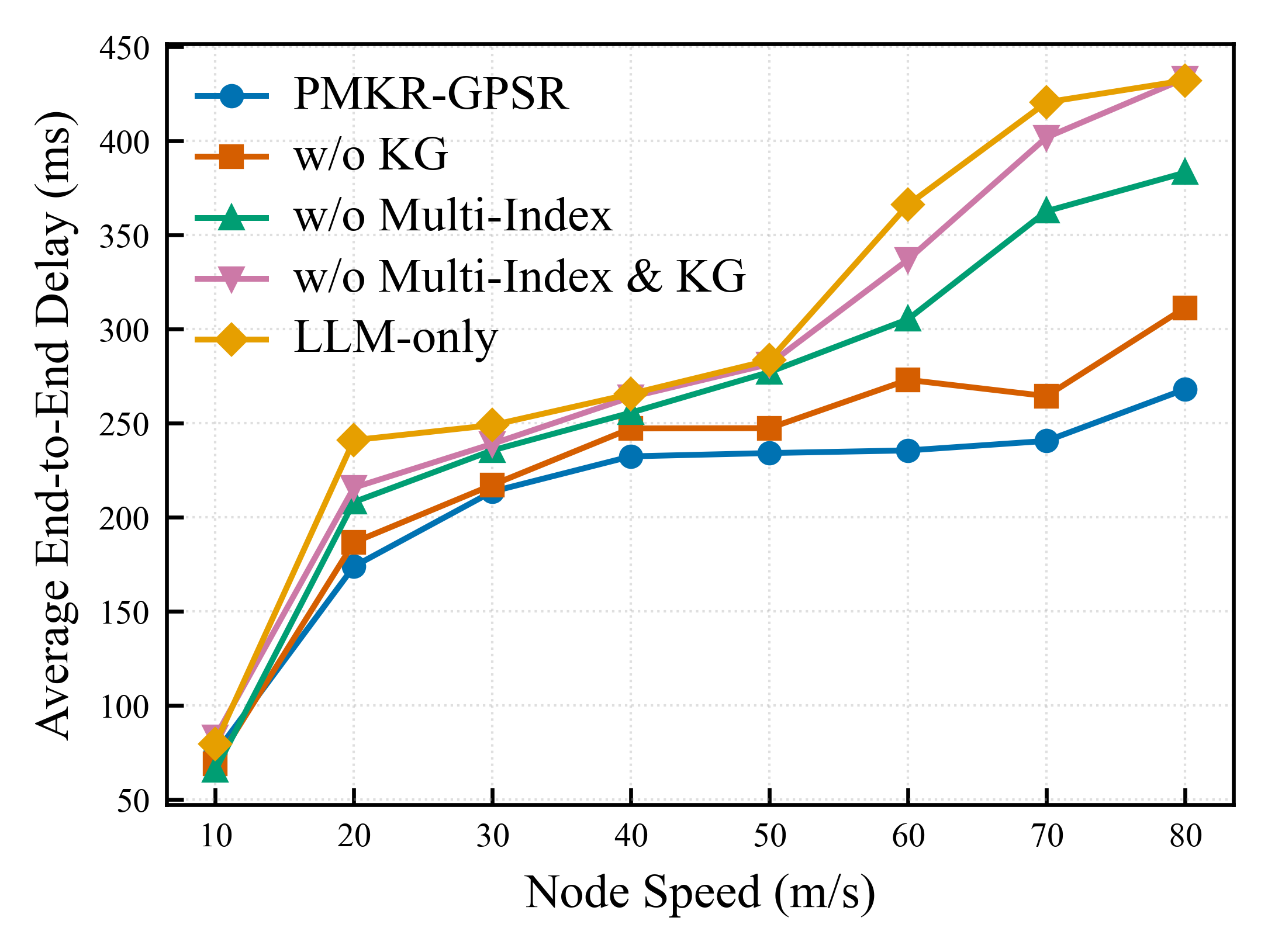}
        \caption{Delay: Ablation.}
        \label{fig:delay_vs_speed_index_ablation}
    \end{minipage}

\vspace{-10pt}
\end{figure*}

Fig.\ref{fig:pdr_vs_speed_baselines} illustrates that the PDR of all protocols decreases as the mobility increases. In particular, when the speed reaches 80 m/s, the PDR of the baseline protocols experiences a significant degradation. In contrast, PMKR-GPSR maintains a relatively stable PDR. This improvement benefits from the parameter-specific retrieval mechanism and knowledge-guided reasoning strategy, which adapt routing parameters to network states and reduce the impact of rapid topology variations. CF-GPSR achieves better stability than GPSR by integrating multiple forwarding metrics for more comprehensive neighbor evaluation. QLGR improves routing decisions through reinforcement learning, but its performance is constrained by the requirements of interaction and policy convergence. GPSR exhibits the most significant performance degradation because it relies primarily on geographic distance and does not consider dynamic link characteristics.

As shown in Fig.\ref{fig:delay_vs_speed_baselines}, the E2E delay of all protocols increases with mobility. Other protocols suffer from more significant degradation due to increased environmental complexity and frequent topology changes. PMKR-GPSR achieves the lowest delay under high mobility scenarios by leveraging knowledge-guided reasoning and parameter-specific retrieval. PMKR-GPSR can identify routing strategies that better match current network states and avoid inappropriate forwarding decisions. Consequently, it reduces route failures, retransmissions, and inefficient forwarding operations under dynamic conditions.

Fig.\ref{fig:pdr_vs_speed_index_ablation} and Fig.\ref{fig:delay_vs_speed_index_ablation} present the PDR and E2E delay performance of different ablation variants, respectively. Compared with LLM-only, w/o Multi-Index \& KG achieves improved performance, indicating that retrieved routing experiences provide useful guidance beyond LLM reasoning alone. The comparison between w/o Multi-Index and PMKR-GPSR demonstrates that parameter-specific retrieval provides more relevant experiences for each routing parameter, reducing irrelevant information interference and improving optimization accuracy. Compared with w/o KG, PMKR-GPSR achieves better performance, indicating that knowledge-guided reasoning incorporates routing constraints into the optimization process and prevents unreasonable parameter configurations. These results validate the necessity of the proposed framework, which integrates parameter-specific retrieval and knowledge-guided reasoning to achieve adaptive routing optimization under highly dynamic network conditions. 

\section{Conclusion }
This paper proposed PMKR-GPSR, an LLM-based framework for GPSR optimization. A parameter-specific multi-index retrieval mechanism was developed to improve the relevance of retrieved routing experiences, and a knowledge-guided constraint graph was introduced to regulate routing parameter optimization. Simulation results demonstrate that PMKR-GPSR achieves higher PDR and lower E2E delay than existing methods under high mobility scenarios. These results indicate that the proposed framework can effectively perceive dynamic network variations and timely adjust routing strategies, select more suitable forwarding schemes in highly dynamic FANET environments. 

\bibliographystyle{IEEEtran}
\bibliography{reference}
\end{document}